\documentclass[12pt]{iopart}

\usepackage{iopams}  
\usepackage{graphicx}
\begin{document}

\title{Residual Kondo effect in quantum dot coupled to half-metallic
       ferromagnets}

\author{Mariusz Krawiec}
\address{Institute of Physics and Nanotechnology Center, 
         M. Curie-Sk\l odowska University, Pl. M. Curie-Sk\l odowskiej 1,
	 20-031 Lublin, Poland}
\ead{krawiec@kft.umcs.lublin.pl}

\begin{abstract}
We study the Kondo effect in a quantum dot coupled to half-metallic 
ferromagnetic electrodes in the regime of strong on-dot correlations. Using the 
equation of motion technique for nonequilibrium Green functions in the slave 
boson representation we show that the Kondo effect is not completely suppressed 
for anti-parallel leads magnetization. In the parallel configuration there is 
no Kondo effect but there is an effect associated with elastic cotunneling 
which in turn leads to similar behavior of the local (on-dot) density of states 
(LDOS) as the usual Kondo effect. Namely, the LDOS shows the temperature 
dependent resonance at the Fermi energy which splits with the bias voltage and 
the magnetic field. Moreover, unlike for non-magnetic or not fully polarized 
ferromagnetic leads the only minority spin electrons can form such resonance in 
the density of states. However, this resonance cannot be observed directly in 
the transport measurements and we give some clues how to identify the effect in 
such systems.
\end{abstract}
%\keywords{proximity effect}

\maketitle

%%%%%%%%%%%%%%%%%%%%%%%%%%%%%%%%%%%%%%%%%%%%%%%%%%%%%%%%%%%%%%%%%%%%%%%%%%%%%%

\section{\label{Introduction} Introduction}

Nowadays spin dependent phenomena play important role in the mesoscopic systems 
as they lead to potential applications in nanotechnology (spintronics)
\cite{Prinz,Awschalom} and quantum computing \cite{Awschalom,Loss}. Moreover 
new transport and thermodynamic phenomena can be observed in spintronic devices 
which are associated with the spin of the electron rather than the charge. 
Those include tunnel magnetoresistance (TMR) in magnetic tunnel junctions 
\cite{Maekawa}, spin dependent Andreev reflections \cite{deJong}, non-monotonic 
behavior of the superconducting transition temperature \cite{Izyumov} and 
spontaneous currents in ferromagnet - superconductor proximity systems 
\cite{MK_1} or Kondo effect \cite{Hewson} in quantum dots (QD) coupled to the 
ferromagnetic leads \cite{Pasupathy}-\cite{Swirkowicz}.

The Kondo effect is a prime example of the many body physics in the quantum dot 
systems, i.e. formation of the many body singlet state by the on-dot spin and 
the conduction electron spins. This state gives rise to the resonance at the 
Fermi energy in the QD density of states and zero-bias maximum in differential 
conductance. The Kondo effect was predicted a long time ago 
\cite{Glazman}-\cite{Kawabata}, extensively studied theoretically 
\cite{Meir}-\cite{MK_2}  and confirmed in series of beautiful 
experiments \cite{Goldhaber}-\cite{Sasaki} in the QD coupled 
to the normal (non-magnetic) leads. 

If the normal leads are replaced by the ferromagnetic ones, the spin degrees of 
freedom start to play significant role in the transport and thermodynamic 
properties of the system, eventually leading to new phenomena. One of such new
effects is the splitting of the Kondo resonance
\cite{Martinek_1,Dong_1,Martinek_2,Choi,Sanchez,Utsumi} due to the spin 
dependent quantum charge fluctuations induced by the tunneling between QD and 
spin polarized leads. Remarkably it is possible to recover the full Kondo 
effect (no splitting) by applying of the external magnetic field 
\cite{Martinek_1,Martinek_2,Sanchez,Utsumi}. The splitting of the Kondo 
resonance strongly depends on the alignment as well as on the magnitude of the 
lead magnetizations. In particular when the magnetizations in both leads point 
in opposite directions (anti-parallel alignment), the full equilibrium Kondo 
effect survives for all values of the polarizations and there is no splitting 
of the zero energy resonance. In the differential conductance however the zero 
bias resonance is getting smaller and smaller as the leads become more 
polarized, finally leading to complete disappearance of the Kondo anomaly. On 
the other hand, in parallel configuration (magnetizations in both leads are 
parallel to each other) the Kondo resonance is split and gets suppressed when 
the magnitude of the polarization is being increased. 

The presence of ferromagnetism in the electrodes can also lead to the quantum 
critical point with non-Fermi liquid behavior. It was shown recently 
\cite{Kirchner} that the competition of spin waves (collective low energy 
excitations in a ferromagnet) and the Kondo effect is responsible for such a 
behavior. In this case the critical Kondo effect manifests itself in a 
fractional power law dependences of the conductance on temperature, and AC 
conductance and thermal noise on frequency $\omega$. Thus the QD system with 
ferromagnetic electrodes can help us to understand the quantum critical 
phenomena in heavy fermions and other correlated electron systems.

In the present paper we show that if the leads are fully polarized, so the 
density of states in the leads is non-zero for one electron spin direction 
only, there is the Kondo effect, provided leads are in anti-parallel 
magnetization configuration. On the other hand in the parallel configuration 
there is no usual Kondo effect but there is an effect associated with elastic
cotunneling which leads to similar behavior of density of states as in usual
Kondo effect. Moreover, this effect occurs only in the minority spin electron 
channel when there is an unpaired spin on the dot and strong on-dot Coulomb
interactions, i.e. when QD is in the Coulomb blockade regime.
 
The paper is organized as follows: in the Sec. \ref{Model} theoretical
description of the QD coupled to the external leads is presented.
Sec. \ref{Processes} is devoted to various elastic and inelastic cotunneling 
processes in the case of half-metallic leads. In the rest of the paper the 
numerical results concerning the density of stets (Sec. \ref{Density}), 
applying of the external magnetic field (Sec. \ref{Compensation}) and the 
transport properties (Sec. \ref{Transport}) are discussed, and finally, some 
conclusions are presented in the Sec. \ref{Conclusions}.

%%%%%%%%%%%%%%%%%%%%%%%%%%%%%%%%%%%%%%%%%%%%%%%%%%%%%%%%%%%%%%%%%%%%%%%%%%%%%%

\section{\label{Model} The Model}

Our system under consideration is represented by the single impurity Anderson
model Hamiltonian in the limit of strong on-dot Coulomb interaction 
($U \rightarrow \infty$) in the slave boson representation where the real 
on-dot electron operator $d_{\sigma}$ is replaced by the product of the boson 
$b$ and the fermion $f_{\sigma}$ operators ($d_{\sigma} = b^+ f_{\sigma}$) 
\cite{Coleman,LeGuillou}:
\begin{eqnarray}
H = \sum_{\lambda {\bf k} \sigma} \epsilon_{\lambda {\bf k} \sigma} 
    c^+_{\lambda {\bf k} \sigma} c_{\lambda {\bf k} \sigma} +
    \sum_{\sigma} \varepsilon_{\sigma} f^+_{\sigma} f_{\sigma} +
    \sum_{\lambda {\bf k}} \left(V_{\lambda {\bf k} \sigma} 
    c^+_{\lambda {\bf k} \sigma} b^+ f_{\sigma} + H. c. \right),
\label{Hamilt}
\end{eqnarray}
where $c_{\lambda {\bf k} \sigma}$ stands for the electrons with the single 
particle energy $\epsilon_{\lambda {\bf k} \sigma}$, the wave vector 
${\bf k}$ and the spin $\sigma$ in the lead $\lambda = {\rm{L,R}}$. 
$\varepsilon_{\sigma}$ denotes the dot energy level and $V_{\lambda {\bf k}}$ 
is the hybridization matrix element between the electrons on the dot and those 
in the leads.

Within the Keldysh formalism \cite{Jauho}, the total current 
$I = \sum_{\sigma} I_{\sigma}$ flowing through the quantum dot is given in the
form:
\begin{eqnarray}
I = \frac{e}{\hbar} \sum_{\sigma} \int d\omega 
\frac{\Gamma_{{\rm{L}} \sigma}(\omega) \Gamma_{{\rm{R}} \sigma}(\omega)}
{\Gamma_{{\rm{L}} \sigma}(\omega) + \Gamma_{{\rm{R}} \sigma}(\omega)} 
[f_{\rm{L}}(\omega) - f_{\rm{R}}(\omega)] \rho_{\sigma}(\omega),
\label{current}
\end{eqnarray}
where we have introduced the elastic rate $\Gamma_{\lambda \sigma}(\omega) = 
\sum_{\bf k} |V_{\lambda {\bf k}}|^2 
\delta(\omega - \epsilon_{\lambda {\bf k} \sigma})$, and 
$\rho_{\sigma}(\omega)$ is the spectral function of the dot retarded Green's 
function $G^r_{\sigma}(\omega)$, calculated within the equation of motion 
technique (EOM) in the slave boson representation \cite{MK_2,MK_3}. 

As is well known the EOM technique is reliable in the high temperature regime, 
however it also qualitatively captures the Kondo physics \cite{MK_2}. Moreover 
the EOM is the one of very few techniques which allows to study nonequilibrium 
properties of the spin polarized QD system.

Within this approach the dot retarded Green's function reads: 
\begin{eqnarray}
G^r_{\sigma}(\omega) = \frac{1 - \langle n_{-\sigma} \rangle}{\omega -
\varepsilon_{\sigma} - \Sigma_{0 \sigma}(\omega) - \Sigma_{I \sigma}(\omega)
+i0^+},
\label{Green}
\end{eqnarray}
with non-interacting ($U = 0$)  
\begin{eqnarray}
\Sigma_{0\sigma}(\omega) = 
\sum_{\lambda {\bf k}} \frac{|V_{\lambda {\bf k}}|^2}
{\omega - \epsilon_{\lambda {\bf k} \sigma}} ,
\label{Sigma_0}
\end{eqnarray}
and interacting self-energy 
\begin{eqnarray}
\Sigma_{I\sigma}(\omega) = \sum_{\lambda {\bf k}} 
\frac{|V_{\lambda {\bf k}}|^2 f_{\lambda}(\epsilon_{\lambda {\bf k} -\sigma})}
{\omega - \epsilon_{\lambda {\bf k} -\sigma} - \varepsilon_{-\sigma} + 
\varepsilon_{\sigma}} ,
\label{Sigma_I}
\end{eqnarray}
which is responsible for the generation of the Kondo effect. 
$\langle n_{\sigma} \rangle$ is the average occupation on the QD, calculated
under nonequilibrium within the standard scheme \cite{MK_2,MK_3}. 

To get the splitting of the Kondo resonance in the presence of the 
ferromagnetic leads we follow Ref. \cite{Martinek_1} and replace 
$\varepsilon_{\sigma}$ on the r.h.s. of the Eq. (\ref{Sigma_I}) by 
$\tilde \varepsilon_{\sigma}$, which is found from the self-consistency 
relation
\begin{eqnarray}
\tilde \varepsilon_{\sigma} = \varepsilon_{\sigma} + 
{\rm Re} [\Sigma_{0\sigma}(\tilde \varepsilon_{\sigma}) + 
\Sigma_{I \sigma}(\tilde \varepsilon_{\sigma})] .
\label{e_sigma}
\end{eqnarray}

In numerical calculations we have chosen 
$\Gamma = \sum_{\lambda \sigma} \Gamma_{\lambda \sigma} = 1$ as an energy unit. 
The magnetization in the lead $\lambda$ is defined as 
$p_{\lambda} = \frac{\Gamma_{\lambda \uparrow} - \Gamma_{\lambda \downarrow}}
{\Gamma_{\lambda \uparrow} + \Gamma_{\lambda \downarrow}}$. For half-metallic
leads (HM) we have 
$\Gamma_{{\rm{L}} \uparrow} = \Gamma_{{\rm{R}} \uparrow} = 0.5$ and 
$\Gamma_{{\rm{L}} \downarrow} = \Gamma_{{\rm{R}} \downarrow} = 0$ in 
the parallel configuration, while 
$\Gamma_{{\rm{L}} \uparrow} = \Gamma_{{\rm{R}} \downarrow} = 0.5$ and 
$\Gamma_{{\rm{L}} \downarrow} = \Gamma_{{\rm{R}} \uparrow} = 0$ in the 
anti-parallel configuration. 

%%%%%%%%%%%%%%%%%%%%%%%%%%%%%%%%%%%%%%%%%%%%%%%%%%%%%%%%%%%%%%%%%%%%%%%%%%%%%%

\section{\label{Processes} Tunneling processes}

Before the presentation of the numerical results let us discuss various 
tunneling processes, associated with the second generation of Green functions
obtained in the EOM procedure. They are the second order processes in the 
hybridization and describe elastic and inelastic cotunneling. The cotunneling 
is a process which leaves the charge on the dot unchanged. Moreover, elastic
cotunneling does not change spin on the dot either, thus it leaves the dot in
its ground state. On the other hand, the inelastic one changes the ground state 
\cite{Averin}.

\begin{figure}[h]
\begin{center}
 \resizebox{0.5\linewidth}{!}{
  \includegraphics{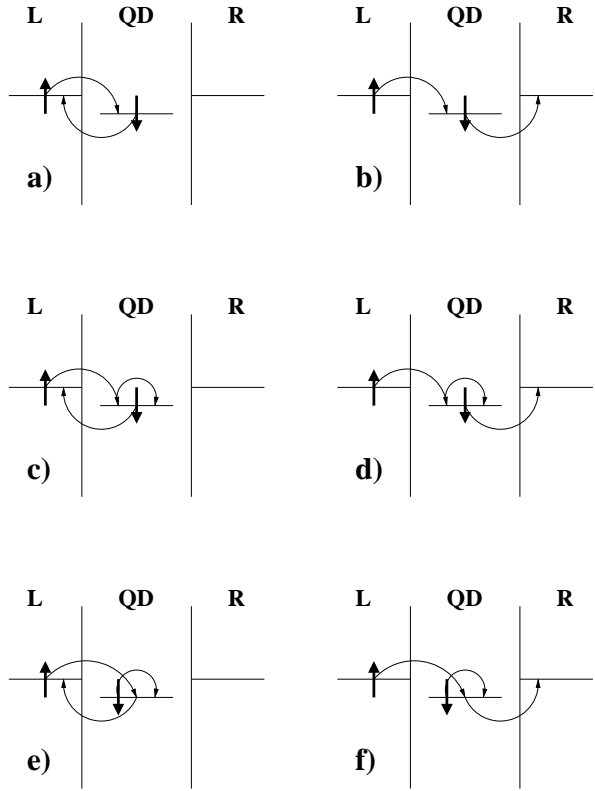}}
\end{center}
 \caption{\label{Fig1} Various tunneling processes associated with second
          generation Green functions in the present EOM approach. For 
	  simplicity we have shown only processes starting from the left 
	  electrode. The remained ones can be obtained by replacing   
	  $L \rightarrow R$ and $\sigma \rightarrow -\sigma$ (see discussion in
	  the text. Left panels show single barrier cotunneling and right 
	  panels - double barrier cotunneling events. Panels a) and b) show 
	  usual inelastic cotunneling, leading to the Kondo effect, changing 
	  the spin on the dot and in one or both electrodes. c) and d) are 
	  similar to a) and b) with additional annihilation and creation of the 
	  spin on the dot. All above processes change spin on the dot and in 
	  the lead(s). The processes displayed in panels e) and f) lead to the 
	  same final spin state on the dot (elastic cotunneling). In general 
	  the only processes contributed to the current across the dot are 
	  associated with double barrier cotunneling, displayed in panels b), 
	  d) and f). In the case of half-metallic ferromagnetic leads in the 
	  anti-parallel configuration the allowed processes are those in b), 
	  d) and e), but they do not contribute to the current, while in 
	  parallel configuration they are those displayed in e) and f) with 
	  the only f) giving a contribution to the current.}
\end{figure}

As is well known the EOM approach is a non-perturbative technique, and one 
cannot assume that those processes are fully taken into account. They are 
rather incorporated in the calculations only qualitatively. It should be 
stressed that they are likely not the only processes as one may get the other 
processes of the same order or contributions to the ones discussed here, while 
going further in EOM procedure, i.e. calculating higher generation GFs. 
Unfortunately, an exact solution of the model is not known, and for the 
present purposes it is enough to consider those, shown in Fig. \ref{Fig1}. 

In Fig. \ref{Fig1} we show such elastic and inelastic single barrier (left
panels) and double barrier (right panels) cotunneling processes which start 
from the spin up electron in the lead L and spin down electron on the dot 
(Coulomb blockade regime). There are also processes which start from lead R and 
can be obtained by replacing $L \rightarrow R$. In general, for not fully 
polarized leads there are processes for opposite spins 
($\sigma \rightarrow -\sigma$) but in the case of half-metallic ferromagnetic
leads in parallel configuration they are not allowed due to lack of the density 
of states for minority spin electrons in the leads. 
We do not show the processes with the same spins on the dot and in the lead, as 
well as those with empty dot state.

Panels a) and b) show the processes in which the electron with spin up
tunnels from the lead L onto the dot and the electron with spin down tunnels
from the dot into the lead L (a)) or R (b)), describing inelastic cotunneling
and leading to the usual Kondo effect, as the spin on the dot is changed. 
Similar situation is displayed in c) and d), namely, the dot starts with spin 
down and ends with spin up. However, during this tunneling event, additionally, 
the spin up electron on the dot is annihilated and created. Those are 
renormalized inelastic cotunneling processes. In all above processes the spin 
on the dot and in the lead(s) is flipped, so those are the Kondo related 
processes. There are two more processes, shown in e) and f), in which there is 
no spin flip. The process starts with spin up in the lead L and ends with the 
same spin in lead L (e)) or R (f)). Similarly, the initial and the final spin 
state on the dot remains the same. Thus they describe also renormalized but 
elastic cotunneling, as the ground state remains unchanged. Both processes 
lead, as we shall see later on, to similar main features of the dot density of 
states as in usual Kondo effect.

As one can read off from Fig. \ref{Fig1}, all the processes are allowed only if 
the leads are not polarized (paramagnetic) or not fully polarized. Moreover, 
the only processes contributed to the current across the dot are double barrier 
cotunneling processes, shown in b), d) and f). 

In the case of the half-metallic ferromagnetic leads in anti-parallel
(AP) configuration, the allowed processes are those shown in b), d) and e). 
However, they do not give any contribution to the current. In fact, processes b)
and d), as they describe double barrier cotunneling, allow for the tunneling 
but once the electron with the spin down tunnels off the dot into the lead R, 
the spin up electron can tunnel from the lead L onto the dot and the further 
tunneling is blocked. The only possibility is the opposite process, namely, the 
spin up electron tunnels from the dot into the lead L and the spin down 
electron from the lead R tunnels onto the dot. Thus, in this case the electron 
transport is completely blocked. The process shown in e) also does not 
contribute to the current as in this case the electron starts and ends in the 
same lead (single barrier cotunneling). As a result, in the case of AP 
configuration there is no current through the quantum dot. 

On the other hand, in the case of the half-metallic leads in parallel (P)
configuration there is one process giving a contribution to the current. This 
is the process shown in Fig. \ref{Fig1}f), coming from the Coulomb interaction. 
In this case the process starts with spin up in the lead L and ends with the 
same spin in the lead R, thus not changing the ground state (double barrier
elastic cotunneling). Note that the inelastic cotunneling is not allowed in 
this case. This process at zero temperature has a finite probability in the 
Coulomb blockade only, i.e. when the dot energy level is below the Fermi energy 
of the electrodes, and there is strong on-dot Coulomb interaction. Moreover, 
this process is allowed only when the dot is occupied by spin down electron. It 
gives a non-zero contribution to the dot density of states below the Fermi 
level (shifted by $\varepsilon_{\uparrow} - \varepsilon_{\downarrow}$ in the 
parallel configuration) only (see Fig. \ref{Fig2} (bottom panel) and Fig. 
\ref{Fig4} (top panel)). It is shown in the next section that this process 
leads to similar main behavior of the density of states of the dot as the usual 
Kondo effect. 

%%%%%%%%%%%%%%%%%%%%%%%%%%%%%%%%%%%%%%%%%%%%%%%%%%%%%%%%%%%%%%%%%%%%%%%%%%%%%%

\section{\label{Density} Density of states}

\begin{figure}[h]
\begin{center}
 \resizebox{0.5\linewidth}{!}{
  \includegraphics{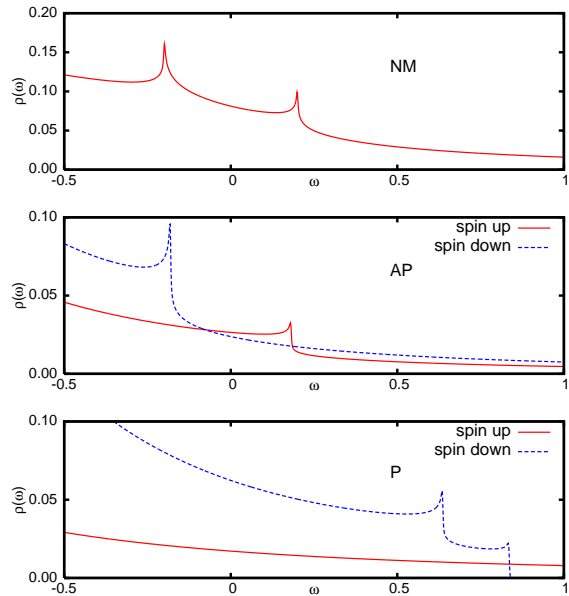}}
\end{center}
 \caption{\label{Fig2} Nonequilibrium ($\mu_{\rm{R}} = - \mu_{\rm{L}} = 0.2$) 
          density of states of the quantum dot coupled to non-magnetic (top 
	  panel), half-metallic leads in anti-parallel configuration (middle 
	  panel) and in parallel configuration (bottom panel). The solid 
	  (dashed) line shows the spin up (down) electron DOS. The model 
	  parameters are: 
	  $\varepsilon_{\uparrow} = \varepsilon_{\downarrow} = -2$, $T=10^{-3}$.
	  All energies are measured in units of $\Gamma$.}
\end{figure}

In the Fig. \ref{Fig2} we show the spin resolved nonequilibrium 
($\mu_{\rm{R}} = - \mu_{\rm{L}} = 0.2$) density of states (DOS) of the 
quantum dot coupled to the external leads. 
Top panel shows the usual DOS of the QD with non-magnetic (NM) leads where two
Abrikosov-Suhl resonances located at the chemical potentials of the leads can 
be observed. The middle panel displays the DOS of the QD coupled to the 
half-metallic electrodes in anti-parallel (AP) configuration 
($p_{\rm{L}} = - p_{\rm{R}} = 1$). It is worthwhile to note that we have 
now spin up Kondo resonance at $\omega = \mu_{\rm{R}}$ (solid line) and no 
resonance at $\omega = \mu_{\rm{L}}$. In this case the spin up on the dot is 
screened by the spins down in the lead R 
($\Gamma_{{\rm{R}} \downarrow} \neq 0$). For spin down electrons situation is 
opposite, the spin down on the dot is screened by the spins up in the lead L 
($\Gamma_{{\rm{L}} \uparrow} \neq 0$) and therefore there is a resonance for 
$\omega = \mu_{\rm{L}}$ and lack of it for $\omega = \mu_{\rm{R}}$. This is 
different from general case of $p < 1$, where two resonances at both chemical 
potentials are present (similarly as in non-magnetic case) \cite{Martinek_1}.

Situation is quite different in the parallel (P) configuration 
($p_{\rm{L}} = p_{\rm{R}} = 1$) where the spin up DOS (solid line in the 
bottom panel of the Fig.\ref{Fig2}) shows no signatures of the Kondo effect as 
the spin up on the dot cannot be screened by spins down in either lead 
($\Gamma_{{\rm{L}} \downarrow} = \Gamma_{{\rm{R}} \downarrow} = 0$). 
However, in the spin down channel the residual Kondo-like state can be produced 
due to the processes shown in the Fig.\ref{Fig1}e) and f). This manifests 
itself in two resonances in the DOS located at 
$\omega = \Delta \varepsilon + \mu_{\lambda}$ (see dashed line in the bottom 
panel of the Fig. \ref{Fig2}), where 
$\Delta \varepsilon = \varepsilon_{\uparrow} - \varepsilon_{\downarrow}$ is the 
splitting due to the ferromagnetic leads. Such splitting has been also observed
in general case of FM ($p < 1$) leads, where the charge fluctuations play
significant role, i.e. when $2 |\varepsilon_d| \neq U$ 
\cite{Martinek_1,Martinek_2,Choi}. Note that in AP configuration there is no 
such splitting, again, in agreement with general case of $p < 1$ polarization 
\cite{Martinek_1,Martinek_2,Choi}. 

Another important effect is a cutoff of the one of the resonances at the energy 
$\omega = \Delta \varepsilon + \mu_R$. This indicates that there is no direct 
tunneling of the spin down electrons and the resulting density of states comes 
form the virtual processes shown in Fig. \ref{Fig1}e) and f) only, as discussed 
in Sec. \ref{Processes}.

\begin{figure}[h]
\begin{center}
 \resizebox{0.5\linewidth}{!}{
  \includegraphics{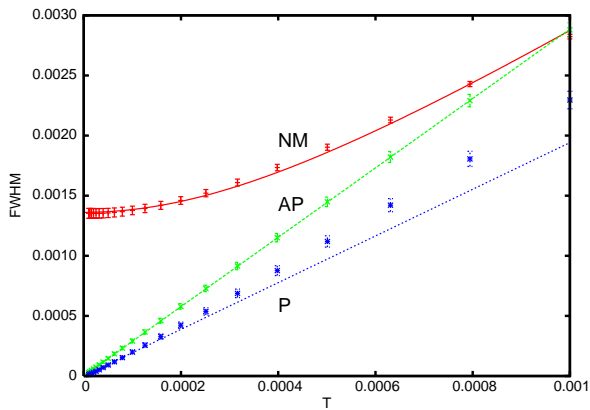}}
\end{center}
 \caption{\label{Fig3} The low temperature behavior of the full width at the 
          half maximum (FWHM) of the narrow (Kondo) resonance in NM, AP and P 
	  configurations.}
\end{figure}

Figure \ref{Fig3} shows the low temperature dependence of the full width at the 
half maximum (FWHM) of the narrow (Kondo) resonance in the density of states 
near $\omega_K$, which equals to the Fermi energy 
($\mu_{\rm{L}} =  \mu_{\rm{R}} = 0$) when QD is coupled to the non-magnetic 
or half-metallic leads in AP configuration and 
$\omega_K = \Delta \varepsilon$ for half-metallic leads in P configuration.
The points with errorbars show numerically found values of the FWHM. The Kondo
temperatures in all cases have been found by fitting the function 
$FWHM = \sqrt{T^2_K + a T^2}$ to those points. In th NM case it gives 
$T_K = 1.36 \cdot 10^{-3}$ while in the AP - $T_K = 1.07 \cdot 10^{-5}$. The 
lower $T_K$ in AP configuration steams from the fact that the Kondo state in 
this case is formed by the electron spins in one lead only for a given 
direction of the spin on the dot. In the P configuration $T_K$ is equal to zero 
as in this case there is no Kondo effect. Moreover in the P configuration the 
FWHM significantly deviates from the linear behavior for higher temperatures. 
On the other hand in the NM and AP configurations it is still linear above 
$T = 10^{-2}$ (not shown in the Fig. \ref{Fig3}).

%%%%%%%%%%%%%%%%%%%%%%%%%%%%%%%%%%%%%%%%%%%%%%%%%%%%%%%%%%%%%%%%%%%%%%%%%%%%%%

\section{\label{Compensation} Compensation by the external magnetic field}

\begin{figure}[h]
\begin{center}
 \resizebox{0.5\linewidth}{!}{
  \includegraphics{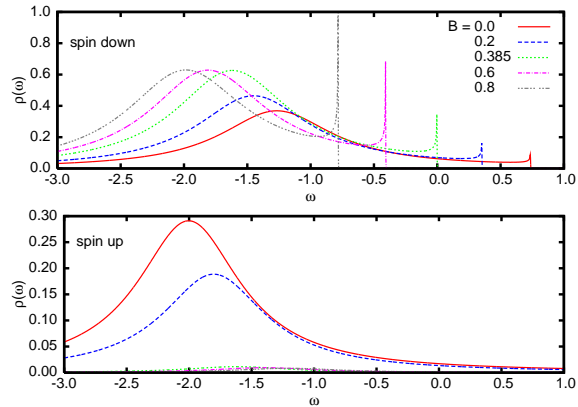}}
\end{center}
 \caption{\label{Fig4} Equilibrium ($\mu_{\rm{L}} =  \mu_{\rm{R}} = 0$) 
          spin down (top panel) and spin up (bottom panel) density of states 
	  for various values of the external magnetic field indicated in the 
	  figure. At $B = B_{\rm{comp}} = 0.385$ there is no splitting of the 
	  dot energy level, thus the Kondo effect is compensated. Note strong 
	  suppression of the spin up density of states for 
	  $B \geq B_{\rm{comp}}$.}
\end{figure}

\begin{figure}[h]
\begin{center}
 \resizebox{0.5\linewidth}{!}{
  \includegraphics{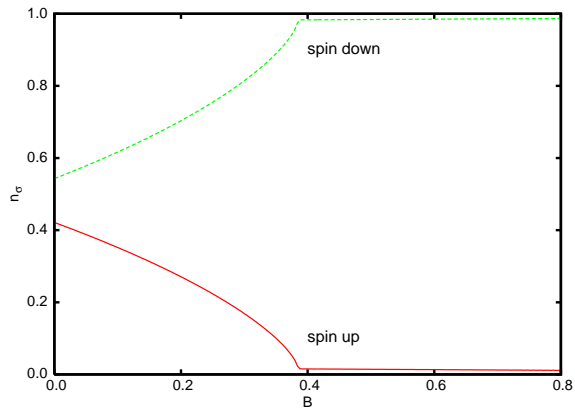}}
\end{center}
 \caption{\label{Fig5} Evolution of the average spin dependent occupation 
          number with the external magnetic field $B$. Note that with 
	  increasing of the $B$ field the spin polarization (difference 
	  between the occupation of the spin up and the spin down electrons) 
	  also increases and remains almost constant (changes very slowly) for 
	  $B > B_{\rm{comp}} = 0.385$.}
\end{figure}

Now, the question arises if the residual Kondo-like effect in P configuration 
can be compensated by the external magnetic field $B$. Compensation in this 
case means no splitting of the dot energy levels $\Delta \varepsilon = 0$. In 
the Fig. \ref{Fig4} we show the equilibrium 
($\mu_{\rm{L}} =  \mu_{\rm{R}} = 0$) spin down (top panel) and spin up 
(bottom panel) density of states for different values of $B$ field. 
As it is evident, magnetic field shifts the spin down Kondo resonance and at 
$B = B_{\rm{comp}} = 0.385$ it reaches the Fermi energy. No splitting 
$\Delta \varepsilon$ is observed in this case. Note that the Kondo resonance 
rapidly grows as $B$ field is increased while the broad resonance around the 
dot energy level initially also grows up but for $B > B_{\rm{comp}}$ remains 
almost unchanged (position of it changes only). The spin up density of states 
(bottom panel) shows different behavior. Namely, the broad charge fluctuation 
resonance starts to decrease with increasing of the $B$ field and at 
$B = B_{\rm{comp}}$ is remarkably small. This effect is associated with the 
change of the average occupation for different spin directions on the dot 
($\langle n_{\sigma} \rangle$). When the $B$ field increases, 
$\varepsilon_{\uparrow}$ moves towards the Fermi energy while 
$\varepsilon_{\downarrow}$ moves in opposite direction and therefore 
$\langle n_{\uparrow} \rangle$ decreases while $\langle n_{\downarrow} \rangle$
increases its value. This is clearly seen in the Fig. \ref{Fig5}, where the
occupations of the dot for both spin directions are shown. 
As one can see the spin up (down) occupation number decreases (increases) with
the increasing of the external magnetic field $B$. As soon as the $B$ field 
exceeds $B_{\rm{comp}} = 0.385$, so one can think in this case about
overcompensated residual Kondo-like effect, both occupation numbers remain 
almost constant, strictly speaking, they change very slowly towards 1 in the 
spin down channel (fully occupied state) and 0 in the spin up channel (empty 
state). 

\begin{figure}[h]
\begin{center}
 \resizebox{0.5\linewidth}{!}{
  \includegraphics{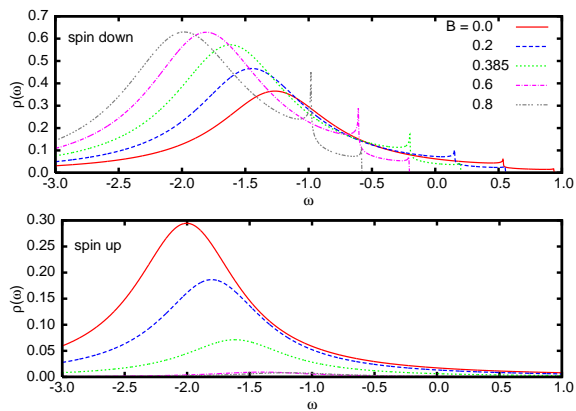}}
\end{center}
 \caption{\label{Fig6} Non-equilibrium 
          ($\mu_{\rm{L}} =  -\mu_{\rm{R}} = -0.2$) spin down (top panel) 
	  and spin up (bottom panel) density of states for the same values of 
	  the external magnetic field as in Fig. \ref{Fig3}.}
\end{figure}

Corresponding non-equilibrium density of states for different magnetic fields 
is shown in Fig. \ref{Fig6}.
While the behavior is similar to that shown in Fig. \ref{Fig3}, except the
fact that there are two resonances now, one can see larger spin up density of 
states at $B = B_{\rm{comp}}$. This is simply due to the smaller value of the
spin down electron occupation number. 

As we have seen the residual Kondo-like effect can be compensated by the 
external magnetic field. However, the compensation in this case means something 
different than in general case of $p < 1$. First of all, we have to remember
that there is only one resonance (for one spin direction only) in the DOS, 
which can be shifted to the Fermi energy by the external magnetic field. This 
is what we call the compensation. In general case of $p < 1$ there are two
resonances (for both spin directions) in the DOS, which can be moved to the 
Fermi energy. Moreover, as it has been shown \cite{Martinek_1,Martinek_2}, the
occupations for both spin directions become equal at $B = B_{\rm{comp}}$ -
there is no spin polarization for such external magnetic field. In the case of
HM ($p = 1$) leads, there is non-zero spin polarization, i.e. 
$\langle n_{\uparrow} \rangle \neq \langle n_{\downarrow} \rangle$ (see the
Fig.\ref{Fig5}). Consequently, the spin on the dot cannot be fully screened by 
the spins in the leads, and there is no Kondo effect in the common sense. This 
is the main and important difference between the compensation effect in the 
case of HM and FM ($p < 1$) leads. 

Finally, we would like to comment on the compensation of the Kondo effect in 
general case of the $p < 1$ within the present approach and compare it to the 
other works known in the literature \cite{Martinek_1,Martinek_2,Choi}. The 
obtained results and the conclusions are qualitatively the same as in the 
papers mentioned above. However, the values of the $B_{\rm{comp}}$ in the 
present work differs from those in Ref. \cite{Martinek_1} due to the fact that 
we have assumed spin dependent bandwidths in electrodes in order to have the 
densities of states in the leads normalized to $1$. For this reason our 
approach gives the values of the $B_{\rm{comp}}$ smaller than those of 
Ref. \cite{Martinek_1}, also obtained within EOM technique. Without this 
normalization requirement we get perfect quantitative agreement with that
approach.

All this shows clear evidence of the usual Kondo effect in the DOS of the 
quantum dot coupled to the half-metallic ferromagnets in AP configuration and
similar cotunneling related effect in P configurations. In both cases the 
density of states shows the splitting of the zero energy resonance caused by 
the nonequilibrium conditions ($\mu_{\rm{L}} \neq \mu_{\rm{R}}$) or the 
exchange field coming from the electrodes in P configuration and finally the 
compensation of it by the external magnetic field. Unfortunately, it is not 
possible to measure directly the density of states in transport experiments and 
the problem arises how to confirm this effect experimentally. In AP 
configuration there is Kondo effect for both spin directions but the tunneling 
current is zero in both cases due to the product of 
$\Gamma_{{\rm{L}} \sigma}(\omega) \Gamma_{{\rm{R}} \sigma}(\omega)$ 
(see Eq.(\ref{current})), which vanishes (see also discussion in the previous
section). On the other hand, in P configuration the cotunneling related effect 
is present for minority spin electrons but there is no density of states in the 
electrodes for this spin direction. Hopefully, in this case presence of the 
effect in the minority electron channel modifies also the transport properties 
in the other channel, therefore in general it is possible to get some 
information on this effect.

%%%%%%%%%%%%%%%%%%%%%%%%%%%%%%%%%%%%%%%%%%%%%%%%%%%%%%%%%%%%%%%%%%%%%%%%%%%%%%

\section{\label{Transport} Transport properties}

\begin{figure}[h]
\begin{center}
 \resizebox{0.5\linewidth}{!}{
  \includegraphics{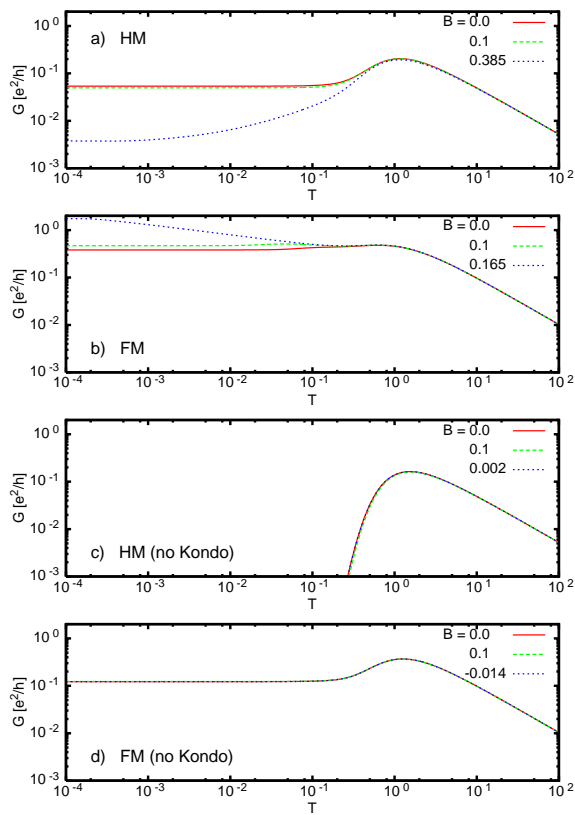}}
\end{center}
 \caption{\label{Fig7} The linear conductance as a function of the temperature 
          for the quantum dot coupled to the half-metallic a) and ferromagnetic 
	  leads b) with polarization $p=0.8$. Note different behavior with 
	  respect to the external magnetic field. For ferromagnetic leads b) it 
	  is possible to get usual Kondo effect by tuning of the magnetic 
	  field. For half-metallic leads applying of the magnetic field 
	  $B = B_{\rm{comp}}$ suppresses the conductance. Panels c) and d) 
	  show the same as a) and b) respectively with neglected cotunneling 
	  like correlations.}
\end{figure}

Figure \ref{Fig7} shows the temperature dependence of the linear conductance
$G = \frac{dI}{dV}|_{V \rightarrow 0}$ for different values of the external 
magnetic field $B$.
At zero magnetic field conductance of QD coupled to the half-metallic leads is 
almost constant at low temperatures (panel a)), similarly as for QD with 
ferromagnetic leads where the polarization is $p=0.8$ (panel b)). When $B$ 
field increases, $G$ of HM starts do decrease, unlike for FM system, where it 
grows up, and finally the Kondo effect is fully compensated at 
$B_{\rm{comp}} = 0.165$. In HM system at $B = B_{\rm{comp}} = 0.385$ (note
that $B_{\rm{comp}}$ are different in HM and FM systems, simply due to
different lead polarizations), the conductance decreases with $T$ and at low 
temperatures is an order of magnitude smaller than at $B = 0$. The decrease of 
$G$ is related to the suppression of the DOS in majority spin channel, as it 
can be read off from the Fig. \ref{Fig4}. Such behavior of the conductance 
remains in agreement with the results obtained within numerical renormalization
group technique (see the Fig. 4b) of Ref. \cite{Martinek_2}), where $G$ is 
plotted as a function of the polarization for $B = B_{\rm{comp}}$. For almost 
all values of the polarizations $p$ the conductance is equal for both spin 
directions, except for $p$ close to 1, where the $G$ becomes spin polarized. In 
our case of $p = 1$, the conductance is fully polarized, as the only one spin 
channel contributes to the transport. 

We have also calculated $G$ in HM (panel c)) and FM (d)) system without
elastic cotunneling and Kondo like correlations taken into account, neglecting 
interacting self-energy $\Sigma_{I\sigma}(\omega)$ in Eq.(\ref{Green}), which 
corresponds to skipping of all tunneling processes shown in the Fig. 
\ref{Fig1}. It is clearly seen that the whole low temperature contribution to 
the conductance of QD with HM leads is due to the renormalized elastic
cotunneling processes and therefore we can conclude that non-zero $G$ is a 
signature of this effect. Conductance of QD coupled to FM leads without
cotunneling correlations (panel d)) shows similar behavior as $G$ for HM leads 
with those processes (panel a)) at $B = 0$. However, lack of the $B$ dependence 
in this case can easily distinguish it from general case of QD coupled to HM 
leads. Therefore, such spectacular behavior of the conductance of the QD with 
HM leads can be, in general, possible to observe in transport measurements. 

\begin{figure}[b]
\begin{center}
 \resizebox{0.5\linewidth}{!}{
  \includegraphics{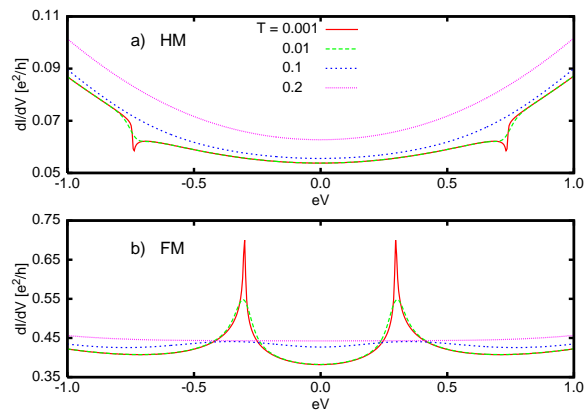}}
\end{center}
 \caption{\label{Fig8} Differential conductance $dI/dV$ of the QD coupled to 
          HM a) and FM b) electrodes for various temperatures.}
\end{figure}

Additional insight into the problem can be reached from the behavior of the 
differential conductance vs bias voltage 
$eV = \mu_{\rm{L}} - \mu_{\rm{R}}$, displayed in the Fig. \ref{Fig8}. 
As one can see in the figure (panel a)) there are temperature dependent 
small kinks at $eV = \pm \Delta \varepsilon$. For comparison the differential 
conductance of QD coupled to FM electrodes with $p = 0.8$ is also shown (panel 
b)). As the temperature grows up, those kinks become suppressed. This is an 
additional clue which can be verified experimentally. We have observed no such 
kinks neither in HM nor in FM system without cotunneling like correlations 
taken into account.

%%%%%%%%%%%%%%%%%%%%%%%%%%%%%%%%%%%%%%%%%%%%%%%%%%%%%%%%%%%%%%%%%%%%%%%%%%%%%%

\section{\label{Conclusions} Conclusions}

In conclusion we have studied the properties of the quantum dot coupled to the
half-metallic leads. In the case of parallel configuration, the effect 
associated with elastic cotunneling, which leads to similar behavior of the
density of states, can be observed. The density of states shows the splitting 
of the zero energy resonance caused by the nonequilibrium conditions 
($\mu_{\rm{L}} \neq \mu_{\rm{R}}$) or the exchange field coming from the 
electrodes and finally the compensation of it by the external magnetic field. 
However the compensation means the shift of the resonance (in the DOS) to the 
Fermi energy only, without additional conditions of equal occupations and equal 
conductances for both spin directions, as in the case of $p < 1$. This effect 
can be observed experimentally measuring the temperature dependence of the 
linear and differential conductances in the external magnetic field. On the 
other hand, in the case of AP configuration, the DOS shows usual Kondo effect 
but the transport is completely suppressed. 

\section*{Acknowledgments}
The author would like to thank Professor K. I. Wysoki\'{n}ski for valuable
discussions and critical reading of the manuscript. 
This work has been supported by the KBN Grant No. 1 P03B 004 28.

%%%%%%%%%%%%%%%%%%%%%%%%%%%%%%%%%%%%%%%%%%%%%%%%%%%%%%%%%%%%%%%%%%%%%%%%%%%%%%


\section*{References}
\begin{thebibliography}{99}
%
\bibitem{Prinz} G. A. Prinz, Science {\bf 282}, 1660 (1998).
%
\bibitem{Awschalom} {\it Semiconductor Spintronics and Quantum Computation}, 
                    ed. D. Awschalom, D. Loss, D. Samarth, Springer, New York 
		    (2002).
%
\bibitem{Loss} D. Loss, D. P. DiVincenzo, Phys. Rev. {\bf A57}, 120 (1998).
%
\bibitem{Maekawa} {\it Spin Dependent Transport in Magnetic Nanostructures}, 
                   ed. S. Maekawa, T. Shinjo, Taylor \& Francis, London and New
		   York (2002).
%
\bibitem{deJong} M. J. M. de Jong, C. W. J. Beenakker, Phys. Rev. Lett. 
                 {\bf 74}, 1657 (1995).
%
\bibitem{Izyumov} Y. A. Izyumov, Y. N. Proshin, M. G. Khusainov, Phys. Usp. 
                  {\bf 45}, 109 (2002).
%
\bibitem{MK_1} M. Krawiec, B. L. Gy\"{o}rffy, J. F. Annett, Phys. Rev. 
               {\bf B66}, 172505 (2002); 
               Eur. Phys. J. {\bf B32}, 163 (2003); 
	       Physica C {\bf 387}, 7 (2003); 
	       Phys. Rev. {\bf B70}, 134519 (2004).
%
\bibitem{Hewson} A. C. Hewson, {\it The Kondo Problem to Heavy Fermions}, 
                 Cambridge University Press, Cambridge, (1993).
%
\bibitem{Pasupathy} A. N. Pasupathy, R. C. Bialczak, J. Martinek, J. E. Grose, 
                    L. A. K. Donev, P. L. McEuen, D. C. Ralph, Science 
		    {\bf 306}, 85 (2004). 
%
\bibitem{Nygard} J. Nygard, W. F. Koehl, N. Mason, L. DiCarlo, C. M. Marcus, 
                 cond-mat/0410467.
%
\bibitem{Sergueev} N. Sergueev, Q. F. Sun, H. Guo, B. G. Wang, J. Wang, Phys. 
                   Rev. {\bf B65}, 165303 (2002).
%
\bibitem{Martinek_1} J. Martinek, Y. Utsumi, H. Imamura, J. Barna\'{s}, 
                     S. Maekawa, J. K\"{o}nig, G. Sch\"{o}n, Phys. Rev. Lett. 
		     {\bf 91}, 127203 (2003).
%
\bibitem{Lu} R. Lu, Z. -R. Liu, cond-mat/0210350.
%
\bibitem{Zhang} P. Zhang, Q. K. Xue, Y. P. Wang, X. C. Xie, Phys. Rev. Lett. 
                {\bf 89}, 286803 (2002).
%
\bibitem{Ma_1} J. Ma, B. Dong, X. L. Lei, Commun. Theor. Phys. {\bf 43}, 341 
               (2005).
%
\bibitem{Bulka} B. R. Bu\l ka, S. Lipi\'{n}ski, Phys. Rev. {\bf B67}, 024404
                (2003).
%
\bibitem{Lopez} R. L\'{o}pez, D. S\'{a}nchez, Phys. Rev. Lett. {\bf 90}, 116602
                (2003).
%
\bibitem{Dong_1} B. Dong, H. L. Cui, S. Y. Liu, X. L. Lei, J. Phys.: Condens. 
                 Matt. {\bf 15}, 8435 (2003).
%
\bibitem{Martinek_2} J. Martinek, M. Sindel, L. Borda, J. Barna\'{s}, 
                     J. K\"{o}nig, G. Sch\"{o}n, J. von Delft, Phys. Rev. Lett. 
		     {\bf 91}, 247202 (2003).
%
\bibitem{Choi} M. -S. Choi, D. Sanchez, R. Lopez, Phys. Rev. Lett. {\bf 92}, 
               056601 (2004).
%
\bibitem{Ma_2} J. Ma, X. L. Lei, Europhys. Lett. {\bf 67}, 432 (2004).
%
\bibitem{Konig} J. K\"{o}nig J. Martinek, J. Barna\'{s}, G. Sch\"{o}n, 
                {\it CFN Lectures on Functional Nanostructures}, Eds. K. Busch 
		et al., Lecture Notes in Physics {\bf 658}, Springer, 145 
		(2005).
%
\bibitem{Martinek_3} J. Martinek, M. Sindel, L. Borda, J. Barna\'{s}, R. Bulla, 
                     J. K\"{o}nig, G. Sch\"{o}n, S. Maekawa, J. von Delft, 
		     cond-mat/0406323.
%
\bibitem{Tanaka} Y. Tanaka, N. Kawakami, J. Phys. Soc. Japan {\bf 73}, 2795
                 (2004).
%
\bibitem{Sanchez} D. Sanchez, R. Lopez, M. -S. Choi, J. Supercond. {\bf 18}, 
                  251 (2005).
%
\bibitem{Utsumi} Y. Utsumi, J. Martinek, G. Sch\"{o}n, H. Imamura, S. Maekawa, 
                 cond-mat/0501172.
%
\bibitem{Swirkowicz} R. \'{S}wirkowicz,  M. Wilczynski, J. Barna\'{s}, 
                     cond-mat/0501605.
%
\bibitem{Glazman} L. I. Glazman, M. E. Raikh, JETP Lett. {\bf 47}, 452 (1988).
%
\bibitem{Ng} T. K. Ng, P. A. Lee, Phys. Rev. Lett. {\bf 61}, 1768 (1988).
%
\bibitem{Kawabata} A. Kawabata, J. Phys. Soc. Japan {\bf 60}, 3222 (1991).
%
\bibitem{Meir} Y. Meir, N. S. Wingreen, P. A. Lee, Phys. Rev. Lett. {\bf 66}, 
               3048 (1991); 
               Phys. Rev. Lett. {\bf 70}, 2601 (1993).
%
\bibitem{Hershfield} S. Hershfield, J. H. Davies, J. W. Wilkins, Phys. Rev. 
                     Lett. {\bf 67}, 3720 (1991); 
		     Phys. Rev. {\bf B46}, 7046 (1992).
%
\bibitem{Wingreen} N. S. Wingreen, Y. Meir, Phys. Rev. {\bf B49}, 11 040 (1994).
%
\bibitem{MK_2} M. Krawiec, K. I. Wysoki\'{n}ski, Phys. Rev. {\bf B66}, 165408
               (2002).
%
\bibitem{Goldhaber} D. Goldhaber-Gordon, H. Shtrikman, D. Mahalu, 
                    D. Abusch-Magder, U. Meirav, M. A. Kastner, Nature 
		    {\bf 391}, 156 (1998).
%
\bibitem{Cronenwett} S. M. Cronenwett, T. H. Oosterkamp, L. P. Kouwenhoven, 
                     Science {\bf 281}, 540 (1998).
%
\bibitem{Schmid} J. Schmid, J. Weis, K. Eberl, K. von Klitzing, Physica B 
                 {\bf 256-258}, 182 (1998); 
                 Phys. Rev. Lett. {\bf 84}, 5824 (2000).
%
\bibitem{Simmel} F. Simmel, R. H. Blick, J. P. Kotthaus, W. Wegscheider, 
                 M. Bichler, Phys. Rev. Lett. {\bf 83}, 804 (1999).
%
\bibitem{Sasaki} S. Sasaki, S. De Franceschi, J. M. Elzerman, 
                 W. G. van der Wiel, M. Eto, S. Tarucha, L. P. Kouwenhoven, 
		 Nature {\bf 405}, 764 (2000).
%
\bibitem{Kirchner} S. Kirchner, L. Zhu, Q. Si, D. Natelson, Proc. Natl. Acad. 
                   Sci. USA {\bf 102}, 18824 (2005).
%
\bibitem{Coleman} P. Coleman, Phys. Rev. {\bf B29}, 3035 (1984).
%
\bibitem{LeGuillou} J. C. Le Giullou, E. Ragoucy, Phys. Rev. {\bf B52}, 2403
                    (1999). 
%
\bibitem{Jauho} H. Haug, A. P. Jauho, {\it Quantum Kinetics in Transport and
                Optics of Semiconductors}, Springer, Berlin (1996).
%
\bibitem{MK_3} M. Krawiec, K. I. Wysoki\'{n}ski, Solid State Commun. {\bf 115},
               141 (2000).
%
\bibitem{Averin} D. V. Averin, Yu. V. Nazarov, in {\it Single Charge 
                 Tunneling}, Eds. H. Grabert, M. H. Devoret, Plenum Press
		 (1992), NATO ASI Series B 294, p. 217.
%
\end{thebibliography}
\end{document}